\documentstyle[aps,prl,epsfig,floats,color]{revtex}

\begin{document}
\draft
\wideabs{

\title{Edge electron states for quasi-one-dimensional organic
   conductors in the magnetic-field-induced spin-density-wave phases}

\author{K. Sengupta, Hyok-Jon Kwon, and Victor M. Yakovenko}

\address{Department of Physics and Center for Superconductivity
   Research, University of Maryland, College Park, MD 20742-4111}

\date{\bf cond-mat/0006050, v.1: June 03, 2000, v.2: January 17, 2001}
% version N

\maketitle

\begin{abstract}%
We develop a microscopic picture of the electron states localized at
the edges perpendicular to the chains in the Bechgaard salts in the
quantum Hall regime.  In a magnetic-field-induced spin-density-wave
state (FISDW) characterized by an integer $N$, there exist $N$
branches of chiral gapless edge excitations.  Localization length is
much longer and velocity much lower for these states than for the edge
states parallel to the chains.  We calculate the contribution of these
states to the specific heat and propose a time-of-flight experiment to
probe the propagating edge modes directly.
\end{abstract}

\pacs {PACS numbers: 74.70Kn, 75.30Fv, 73.40Hm, 73.20-r}
}

Quasi-one-dimensional (Q1D) organic conductors of the $\rm(TMTSF)_2X$
family \cite{TMTSF} (also known as the Bechgaard salts) are highly
anisotropic crystals that consist of parallel conducting chains.  The
electron transfer integrals along the chains (in the {\bf a}
direction) and transverse to the chains (in the {\bf b} and {\bf c}
directions) are $t_a=250$ meV, $t_b=25$ meV, and $t_c=1.5$ meV
\cite{Yamaji}.  Because of the strong Q1D anisotropy, the Fermi
surfaces of these materials are open and consist of two disconnected
sheets located near $\pm k_F$, which are the Fermi momenta along the
chains.  In the presence of a strong magnetic field along the {\bf c}
axis, the interplay between the nesting property of the open Fermi
surface and the quantization of electron orbits due to the magnetic
field leads to a cascade of the magnetic-field-induced
spin-density-wave (FISDW) phase transitions \cite{Yamaji,Schegolev}.
FISDW opens an energy gap $\Delta_N$ in the electron spectrum at the
Fermi level.  Within each FISDW phase, the Hall conductivity per {\bf
a}-{\bf b} layer per electron spin is quantized at zero temperature as
$\sigma_{xy}=Ne^2/h$, where $N$ is an integer that characterizes the
FISDW, $e$ is the electron charge, and $h$ is the Planck constant.

The theory of the quantum Hall effect (QHE) in the Bechgaard salts was
mostly oriented toward the bulk description \cite{Schegolev}.
Particularly, the (2+1)D Chern-Simons term has been derived
microscopically for the bulk effective action of the system in Ref.\
\cite{Yakovenko98}.  According to general principles
\cite{Halperin82,Kane97}, a system where the Hall effect is quantized
with an integer number $N$ should have $N$ gapless chiral electron
states at the edges of the sample.  However, specific microscopic
realization of such edge modes in Q1D organic conductors was not
clear.  In these systems, the electron states at the edges parallel
and perpendicular to the conducting chains must be very different
because of the strong anisotropy.  The edges parallel to the chains
were studied microscopically in Ref.\ \cite{Yakovenko96}.  It was
shown, that the $+k_F$ electrons on the first $N$ chains at one edge
and the $-k_F$ electrons on the last $N$ chains at the other edge
remain ungapped in the FISDW state.  The velocity of these chiral
modes is equal to the Fermi velocity $v_F$ along the chains.  However,
the edges perpendicular to the chains were not discussed in Ref.\
\cite{Yakovenko96}.  Phenomenological and numerical studies of
multilayered quantum Hall systems \cite{Chalker95,Balents96} did not
address the microscopics and anisotropy of the edge states in Q1D
case.

In this letter, we develop a microscopic theory of the chiral states
in Q1D conductors at the edges perpendicular to the chains.  We find
$N$ branches of gapless edge excitations that are localized along the
chains within the coherence length $\xi_N=\hbar v_F/\Delta_N$.  Their
group velocity perpendicular to the chains is
$v_{\perp}=N\Delta_Nb/\hbar$, where $b$ in the interchain distance in
the {\bf b} direction.  Thus, their localization length is much longer
and the velocity much lower than the corresponding parameters $Nb$ and
$v_F$ for the edge states parallel to the chains.

To explain the origin of the edge states \cite{Davison}, it is useful
first to consider a one-dimensional (1D) charge- or spin-density-wave
(CDW/SDW) system. The mean-field Hamiltonian for the system can be
written as \FL
\begin{eqnarray}
{\mathcal H} &=&\! \int dx \,\psi^{\dagger}(x) 
\Big(-\frac{\hbar^2 \partial_x^2}{2m} + 2 \Delta \cos(2 k_F x + \theta
) \Big ) \psi(x), 
\label{H_1+1}
\end{eqnarray}
where we have chosen the $x$ axis along the chain.  $\psi(x)$ is the
fermion (electron) field, $m$ is the effective mass of the fermions,
$2\Delta\cos(2k_Fx+\theta)$ is the density-wave potential, and
$\theta$ is the phase specifying the density-wave position.  To
simplify the presentation, we suppress the spin structure of the
fermion field and the order parameter, which is not essential for our
purposes.  Following the standard procedure, we introduce a doublet of
fermion fields, $\psi_+$ and $\psi_-$, with momenta close to the Fermi
points $\pm k_F$:
\begin{equation}
\psi(x) = \psi_+(x)\,e^{i k_F x} + \psi_-(x)\,e^{-i k_F x},
\label{doublet}
\end{equation}
and linearize the energy dispersion relation near the Fermi points.
The Hamiltonian can then be written in terms of the fermion doublet
$\psi=(\psi_+,\psi_-)$ as \FL
\begin{eqnarray}
{\mathcal H} &=& \!\int\! dx \,\psi^{\dagger}(x) ( \tau_0
\epsilon_F - \tau_z i \hbar v_F \partial_x  + \tau_x 
\Delta e^{-i \tau_z \theta} ) \psi(x),
\label{onedh}
\end{eqnarray}
where $v_F = \hbar k_F /m$ is the Fermi velocity, $\epsilon_F =
\hbar^2 k_F^2/2m$ is the Fermi energy, and $\tau_x$, $\tau_y$,
$\tau_z$ and $\tau_0$ are the $2\times2$ Pauli and unit matrices
acting on the fermion fields.  From Hamiltonian (\ref{onedh}), one
obtains the eigenvalue equation for the quasiparticles:
\begin{equation}
\left(-\tau_z i \hbar v_F \partial_x  + \tau_x 
\Delta e^{-i \tau_z \theta}\right) \psi = E \psi,
\label{BDg}
\end{equation} 
where $E$ is the quasiparticle energy as measured from the Fermi
energy. For an infinite sample, Eq.\ (\ref{BDg}) has well-known
plane-wave solutions $\psi\propto e^{ikx}$ with the spectrum
\begin{equation}
E_k = \pm\sqrt{\hbar^2 v_F ^2 k^2 + \Delta^2}.
\label{E_k}
\end{equation}
The spectrum (\ref{E_k}) forbids electron states within the energy gap
$\pm\Delta$.

Now let us consider a semi-infinite CDW/SDW system occupying the
positive semispace $x>0$ with an edge at $x = 0$. In this case, the
wave function must vanish at the edge: $\psi(x=0)=0$, or equivalently
$\psi_+(x=0)=-\psi_-(x=0)$. With this boundary condition, in addition
to the delocalized states (\ref{E_k}), Eq.\ (\ref{BDg}) also admits a
localized state with the energy $|E|<\Delta$ inside the gap:
\begin{equation}
\psi_{\pm} = \pm e^{-\kappa x}/\sqrt\kappa, \quad
E = -\Delta \cos\theta, \quad
\kappa = -\sin\theta/\xi.
\label{onedr}
\end{equation}
The wave function (\ref{onedr}) decays in the bulk at a length of the
order of the coherence length $\xi=\hbar v_F/\Delta$.  Eq.\
(\ref{onedr}) is meaningful only when $\kappa>0$, thus the localized
state exists only for $\pi<\theta<2\pi$.  If the CDW/SDW is displaced
by one period, the phase $\theta$ changes from $0$ to $2\pi$, and the
energy of the edge state (\ref{onedr}) changes from $\Delta$ to
$-\Delta$.  This process is the spectral flow \cite{Stone96} from the
upper to the lower energy band (\ref{E_k}).  (The opposite spectral
flow takes place at the other end of the sample.)  The edge state
(\ref{onedr}) is similar to the electron state localized at the kink
soliton in an infinite CDW/SDW system \cite{Brazovskii}.

Now let us consider a Q1D FISDW system.  The Hamiltonian of this
system differs from Hamiltonian (\ref{H_1+1}) by the interchain
tunneling term with the amplitude $t_b$ \cite{Schegolev,Yakovenko98}:
\begin{eqnarray}
{\mathcal H}' &=& \int dx\,\frac{dk_y}{2\pi}\,
\psi^{\dagger}(x,k_y) \Big( -\frac{\hbar^2\partial_x^2}{2m} 
+ 2 t_b \cos (k_y b - G x) \nonumber\\
&& {}+ 2 \Delta \cos(Q_x x + \theta ) \Big) \psi(x,k_y),
\label{H_2+1}
\end{eqnarray}
where $k_y$ is the electron wave vector transverse to the chains.  In
the presence of an external magnetic field $H$, the interchain
tunneling term contains the magnetic wave vector $G=ebH/{\hbar}c$ ($c$
is the speed of light) because of the Peierls-Onsager substitution
$k_y-eA_y/c\hbar$ with the vector potential $A_y=Hx$ in the Landau
gauge.  Another difference between Hamiltonians (\ref{H_1+1}) and
(\ref{H_2+1}) is that the FISDW wave vector $Q_x$ deviates from $2k_F$
by an integer multiple $N$ of $G$: $Q_x=2k_F-NG$
\cite{Schegolev,Yakovenko98}.  To simplify presentation, here we
neglect the coupling $t_c$ between the ({\bf a}-{\bf b}) layers and
model the system as a set of uncoupled 2D planes.  Further, we only
consider the electron tunneling between the nearest-neighboring chains
and also assume that the transverse component of the FISDW wave vector
is zero.  While these assumptions are too simplistic for computation
of the quantities such as $N$, $\Delta$, and $T_c$, they do not affect
our results about the edge states.

Next we introduce the doublet of fermion fields $\psi=(\psi_+,\psi_-)$
as in Eq.\ (\ref{doublet}) and linearize the energy dispersion near
$\pm k_F$.  The Hamiltonian becomes
\begin{eqnarray} 
&&{\mathcal H}' = \int dx \, \frac{dk_y}{2\pi} \,
\psi^{\dagger}(x,k_y) \Big( \tau_0 \epsilon_F - \tau_z i \hbar v_F 
\partial_x \nonumber\\
&& {}+\tau_0 2 t_b \cos(k_y b - G x)
+ \tau_x \Delta e^{i \tau_z (N G x -\theta)} \Big) 
\psi(x,k_y).
\label{twodh}
\end{eqnarray}
Then we make a chiral transformation of the doublet $\psi$:
\begin{eqnarray}
\psi(x,k_y) &=& e^{i\tau_z\varphi(x,k_y)}\psi'(x,k_y), 
\\
\varphi(x,k_y) &=& (2t_b/\hbar\omega_c) \sin(k_y b - G x),
\end{eqnarray}
where $\hbar\omega_c=\hbar v_F G$ is the cyclotron energy.  In terms
of the new field $\psi'$, the Hamiltonian can be written as
\begin{eqnarray} 
{\mathcal H}' &=& \int dx\,\frac{dk_y}{2\pi}\,
\psi'^{\dagger}(x,k_y)(\tau_0\epsilon_F - \tau_z i\hbar v_F \partial_x
\nonumber\\
&& {}+ \tau_x \Delta e^{i\tau_z[NGx-\theta+2\varphi(x,k_y)]}) \psi'(x,k_y).
\label{twodh1}
\end{eqnarray}
The transverse hopping term is now transferred to the phase of the
FISDW order parameter.  Next, we expand that phase factor into a
Fourier series
\begin{eqnarray}
e^{2i\tau_z\varphi(x,k_y)} &=& \sum_n J_n(4t_b/\hbar\omega_c)
e^{i \tau_z n (k_y b - G x)},
\label{e^phi}
\end{eqnarray}
where $J_n$ is the Bessel function of the order $n$.  Substituting
Eq.\ (\ref{e^phi}) into Eq.\ (\ref{twodh1}), we only retain the term
with $n = N$ in the series, since only this term does not have an
oscillatory dependence on $x$ and opens a gap at the Fermi level.
This is the so-called single-gap approximation
\cite{Montambaux86,1gap}.  The Hamiltonian then becomes
\begin{eqnarray} 
{\mathcal H}' &=& \int dx \, \frac{dk_y}{2\pi} \,
\psi'^{\dagger}(x,k_y) \Big( \tau_0 \epsilon_F - \tau_z i \hbar v_F 
\partial_x \nonumber\\
&& {}+ \tau_x {\Delta}_N 
e^{ i \tau_z \left(N k_y b -\theta 
\right) } \Big) \psi'(x,k_y),
\label{twodh2}
\end{eqnarray}
where ${\Delta}_N = \Delta J_N(4t_b/\hbar\omega_c)$ is the
modified gap amplitude. Comparing Eqs.\ (\ref{twodh2}) and
(\ref{onedh}), we see that the Q1D FISDW Hamiltonian has been mapped
onto an effective 1D CDW/SDW Hamiltonian with the gap amplitude
${\Delta}_N$ and the order-parameter phase $\theta - N k_y b$.

\begin{figure}
\centerline{\psfig{file=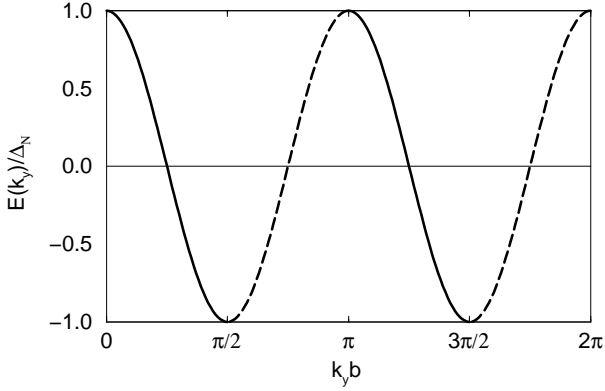,width=\linewidth,angle=0}}
\caption{Energy dispersion of the electron states localized at the
left (solid lines) and right (dashed lines) edges of the sample as a
function of the transverse momentum $k_y$ for $N=2$.}
\label{Graph1}
\end{figure}

\begin{figure}
\centerline{\psfig{file=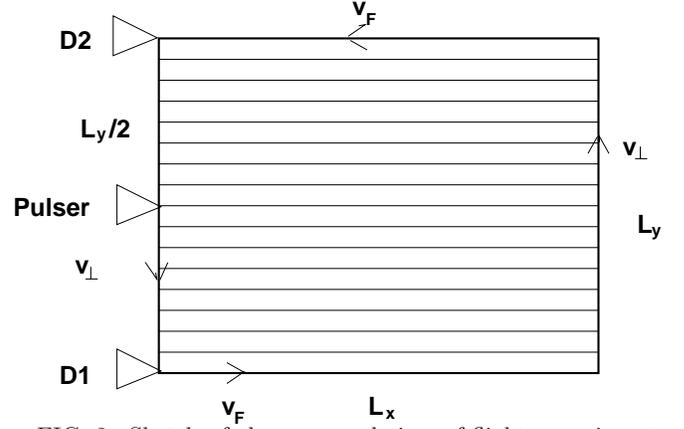,width=\linewidth,angle=0}}
\caption{Sketch of the proposed time-of-flight experiment. The arrows
indicate the directions of the edge modes velocities $v_{\perp}$ and
$v_F$.  The thin lines indicate the conducting chains in the sample.
The pulser sends a pulse, which is detected at different times $t$ and
$t'$ by the detectors D1 and D2.}
\label{Graph2}
\end{figure}

Now let us consider a semi-infinite FISDW system occupying the
positive semispace $x>0$ along the chains and infinite space in the
$y$-direction transverse to the chains.  The wave function $\psi'$
must vanish at the edge: $\psi'_+(0,y)=-\psi'_-(0,y)$.  Following the
same calculation as for the 1D CDW/SDW, we find the edge states with
the energies $|E|<\Delta_N$ inside the gap:
\begin{eqnarray}
&& \psi'_{\pm}(x,k_y) = \pm e^{ik_yy -\kappa x}/\sqrt\kappa,
\quad \kappa =\sin(Nk_yb)/\xi_N,
\label{psi'} \\
&& E_N(k_y) = -\Delta_N\cos(Nk_yb),\quad\xi_N=\hbar v_F/\Delta_N.
\label{twodr}
\end{eqnarray}
Here $E_N(k_y)$ is the quasi-particle energy measured from the Fermi
energy, and we have dropped the now unimportant phase $\theta$.  The
single bound state (\ref{onedr}) is now replaced by the band
(\ref{twodr}) of the edge states labeled by the wave vector $k_y$
perpendicular to the chains.  These states (\ref{psi'}) are bound
along the direction of the chains, but are extended in the direction
transverse to the chains.  At the left edge of the sample, we require
that $\kappa>0$ and $0<Nk_yb<\pi$, and find $N$ branches of the edge
states in the transverse Brillouin zone $0<k_yb<2\pi$.  Complementary
$N$ branches of the edge states determined by the condition $\kappa<0$
and $\pi<Nk_yb<2\pi$ exist at the right edge.  The solid and dashed
lines in Fig.\ \ref{Graph1} show the energy dispersions of the states
localized at the left and right edges, respectively (for $N=2$).  It
is clear that the group velocities of the edge modes $\partial
E_N(k_y)/\hbar\partial k_y$ have opposite signs for the left and right
edges.  Thus, they carry a surface current around the sample, as
indicated by the arrows in Fig.\ \ref{Graph2}.  The sense of
circulation is determined by the sign of $N$, which is controlled by
the sign of the magnetic field $H$.

An effective action for the low-lying edge excitations can be obtained
by linearizing the energy dispersion (\ref{twodr}) near the Fermi
energy.  Since there are $N$ branches of these excitations, the
effective action can be written in terms of an $N$-component chiral
fermion field $\phi=(\phi_1,\phi_2,\ldots,\phi_N)$, as in Ref.\
\cite{Kane97}:
\begin{eqnarray}
S_{\rm edge} &=& \int dt\,dy\,\phi^{\dagger}(t,y) 
(i\hbar\partial_t + i\hbar v_{\perp}\partial_y)\phi(t,y),
\nonumber\\
v_{\perp} &=& \frac{1}{\hbar} \left(
\frac{\partial E_N(k_y)}{\partial k_y}\right)_{E_N(k_y) = 0}
= \frac{N \Delta_N b}{\hbar}, 
\label{edac}
\end{eqnarray}
where $t$ is time.  The group velocity $v_{\perp}$ (\ref{edac}) is
same for all branches.  The velocity is rather low, because it is
proportional to the small FISDW gap ${\Delta}_N$.  The activation
energy $2\Delta_N=6$K was measured in $\rm(TMTSF)_2ClO_4$
\cite{Chamberlin} at $H=25$ T, where $N=1$.  The BCS relation
$\Delta=1.76\:T_c$ gives even bigger value $\Delta_N=5.3$ K for
$T_c=3$ K at $H=9$ T in $\rm(TMTSF)_2ClO_4$ \cite{Scheven95}.  Using
the value $\Delta_N=3$ K for $N=1$, and the lattice spacing $b=0.77$
nm, we find $v_{\perp}=300$ m/s.  The group velocity for the edges
perpendicular to the chains, $v_{\perp}$, is three orders of magnitude
lower than for the edges parallel to the chains, $v_F=190$ km/s (see
Fig.\ \ref{Graph2}).  Nevertheless, the total edge current $I$ is the
same: $I=(Ne/\pi\hbar)\int_0^{\pi/2bN}(\partial E_N/\partial
k_y)\,dk_y=Ne\Delta_N/\pi\hbar=ev_{\perp}/\pi b=20$ nA.  For the
typical sample dimensions $L_x=2$ mm and $L_y=L_z=0.2$ mm and the
interlayer spacing 1.35 nm, the edge current $I$ produces the magnetic
moment $1.2$ nA\,m$^2$ and magnetization $15$ A/m.  However, there are
additional contributions to the total magnetization coming from the
edge states inside the energy gaps opened by the neglected terms in
Eq.\ (\ref{e^phi}) below the Fermi level \cite{1gap}.  Magnetization
has been calculated using the bulk free energy in Ref.\ 
\cite{Montambaux88} and measured experimentally in $\rm(TMTSF)_2ClO_4$
in Ref.\ \cite{Naughton}.

The gapless edge excitations make a contribution $C_e$ to the specific
heat per layer proportional to temperature $T$ at $T\ll\Delta_N$:
\begin{equation}
\frac{C_e}{T} = \frac{N\pi}{3\,\hbar} \left(\frac{2
L_y}{v_{\perp}} + \frac{2 L_x}{v_F} \right)
\approx\frac{2\pi}{3\Delta_N}\,\frac{L_y}{b}.
\label{C/T}
\end{equation}
Using the numbers quoted above, we find that $t_x=L_x/v_F=10$ ns and
$t_y=L_y/v_{\perp}=0.67$ $\mu$s for $N=1$.  Thus, the dominant
contribution to the specific heat comes from the edges perpendicular
to the chains.  According to Eq.\ (\ref{C/T}), the specific heat must
be discontinuous at the boundaries between the FISDW phases, where
$\Delta_N$ changes discontinuously \cite{Montambaux88}.  Note that $N$
cancels out in Eq.\ (\ref{C/T}).  This result is in contrast with the
phenomenological model of Ref.\ \cite{Balents96}, where the edge
velocity was assumed to be the same for all FISDW phases and much
stronger discontinuities $C_e/T\sim N$ were predicted.

The bulk specific heat in the normal state is also proportional to
temperature: $C_b^{\rm(n)}/T=\pi L_xL_y/3\hbar bv_F$.  Its ratio to
the edge specific heat (\ref{C/T}) is roughly equal to the ratio of
the volumes occupied by the bulk and edge states:
$C_b^{\rm(n)}/C_e=L_x/2\xi_N=2.1\times10^3$.  In the FISDW phase, the
bulk specific heat $C_b^{\rm(F)}/T\approx
\sqrt{18/\pi^3}\,(\Delta_N/T)^{5/2}\exp(-\Delta_N/T)\,C_b^{\rm(n)}/T$
is smaller than the edge one at the sufficiently low temperature
$T\le\Delta_N/14$.  Such a regime could have possibly been achieved at
$H=9$ T in the experimental measurements of the specific heat
\cite{Scheven95} performed at $T=0.32$ K in $\rm(TMTSF)_2ClO_4$.  We
estimate the absolute values of $C_e/T$ as $2.2 \times 10^{-3}$
mJ/(K$^2$\,mole) and $C_b^{\rm(n)}/T$ as 4.4 mJ/(K$^2$\,mole).  The
latter estimate roughly matches the experimental value $5$
mJ/(K$^2$\,mole) \cite{Scheven95}.

The edge states picture was confirmed in a multilayered GaAs system by
demonstrating that the conductance $G_{zz}$ perpendicular to the
layers is proportional to the number of the edge states $N$ and the
perimeter of the sample \cite{Druist}.  A similar experiment on the
Bechgaard salts produced inconclusive results \cite{Uji}.  As a
definitive proof of the edge states picture for a FISDW, we propose a
time-of-flight experiment analogous to those performed on GaAs in
Ref.\ \cite{Ashoori92}.  The experimental setup is sketched in Fig.\
\ref{Graph2}.  A pulse is applied by the pulser at the center of the
edge perpendicular to the chains.  The pulse can be electric, as in
Ref.\ \cite{Ashoori92}, and produce a charge perturbation in the
occupation number of the edge states.  Alternatively, the pulse can be
thermal and perturb the Fermi distribution function of the edge
electrons, or magnetic and change the spin population of the edge
states \cite{Awschalom}.  The pulse will travel with the slow velocity
$v_{\perp}$ along the edges perpendicular to the chains and the fast
velocity $v_F$ parallel to the chains.  The pulse will reach the
detector D1 at a time $t =L_y/2v_{\perp}$ and other detector D2 at a
time $t'=3t + 2L_x/v_F\approx3t$.  The difference between $t$ and $t'$
is a signature of the broken time-reversal symmetry in the system.
The inverse delay time $1/t$ is proportional to $v_{\perp}$, given by
Eq.\ (\ref{edac}), and has discontinuities at the FISDW phase
boundaries due to discontinuity of both $N$ and $\Delta_N$.  The pulse
must be shorter than $t=0.33$ $\mu$s for clear resolution and longer
than $\hbar/\Delta_N=2.6$ ps, so that only the low-lying excitations
are probed.

In conclusion, we have shown, starting from a microscopic Hamiltonian,
that there are $N$ chiral gapless electron states bound to the edges
perpendicular to the chains in a FISDW phase characterized by an
integer $N$.  The gapless modes contribute a term linear in
temperature, which dominates the specific heat at low temperatures.
The group velocity of these excitations is rather low:
$v_{\perp}=N\Delta_Nb/\hbar\approx300$ m/s.  We have proposed a
time-of-flight experiment to probe the propagating edge modes
directly.  The chiral character of these states makes them similar to
the edge states in superconductors with broken time-reversal symmetry,
such as $d+is$ and $d_{x^2-y^2}+id_{xy}$ singlet superconductors and
$p_x+ip_y$ triplet superconductors $\rm Sr_2RuO_4$ and 2D $^3$He-A
\cite{Senthil}.  $\rm(TMTSF)_2X$ in the superconducting state is
expected to have nonchiral midgap edge states \cite{midgap}.

%\vspace{-1.5\baselineskip}

\end{document}